\documentstyle[12pt]{article}
\def\beginapjbib{\begingroup \section*{\large \bf References}
         \parskip=.5ex plus 1.0pt
         \def\bibitem{\par \noindent \hangindent\parindent
                \hangafter=1}}
\def\endapjbib{\par \endgroup}
 \def\beq{\begin{equation}}
\def\eeq{\end{equation}}

\begin{document}
\begin{titlepage}
\pagestyle{empty}
\baselineskip=21pt
\begin{flushright}
UMN-TH-1203/93\\
May 1993\\
\end{flushright}
\vskip 0.5in

\begin{center}
{\large{\bf Neutrino Process Nucleosynthesis and \\
 the $^{11}$B/$^{10}$B Ratio}}
\end{center}
\begin{center}
\vskip 0.3in
{ Keith~A.~Olive,$^{1,2}$, Nikos Prantzos,$^{2,3}$ \\
 Sean Scully,$^1$  and Elisabeth Vangioni-Flam,$^2$
}\\

\vskip 0.25in
{\it

$^1${School of Physics and Astronomy,
University of Minnesota, Minneapolis, MN 55455, USA}\\

$^2${Institut d'Astrophysique de Paris, 98bis Boulevard
Arago, 75014 Paris, France}\\

$^3${Service d'Astrophysique, Centre d'Etudes de
Saclay, 91191 Gif sur Yvette, France}
}\\
\vskip 0.5in
{\bf Abstract}
\end{center}
 \baselineskip=18pt

We consider the evolution of the light elements ($Li, Be$ and $B$)
incorporating the effects of their production by both
neutrino process and cosmic-ray nucleosynthesis. We test the viability
of the neutrino process to resolve the long standing problem of the
$^{11}$B/$^{10}$B isotopic ratio which amounts to 4 at the time
of the formation of the solar
system. This hypothesis may be ultimately
constrained by the $B/Be$ ratio observed in halo stars.
Though we are able to obtain a solar isotopic ratio
$^{11}B/^{10}B \simeq 4$, the current paucity of data at low
metallicity prevents us from
making a definitive conclusion regarding the resolution of this
problem.
We show however, that neutrino process nucleosynthesis
leads to a relatively model independent prediction that the $B/Be$
elemental ratio is large ($>$ 50) at low metallicities
($[Fe/H] < -3.0$),
if $Be$ is produced as a secondary element (as is the case
in the conventional
scenario of galactic cosmic-ray nucleosynthesis).
\end{titlepage}
\baselineskip=18pt
{\newcommand{\la}{~\mbox{\raisebox{-.6ex}{$\stackrel{<}{\sim}$}}~}
{\newcommand{\ga}{~\mbox{\raisebox{-.6ex}{$\stackrel{>}{\sim}$}}~}

\section{Introduction}

One of the successes of the standard big bang model is the
prediction of the
primordial abundances of the light nuclei,
$D$, $^{3}He$, $^{4}He$, and $^{7}Li$.
In the standard model, the abundances of these isotopes, which span
ten orders of magnitude (by number), are in accordance with
observations for a narrow range in the baryon density or equivalently
a baryon to photon ratio of $2.8-4 \times 10^{-10}$
(Walker et al. 1991) .

The other light nuclei, $^{6}Li$, $^{9}Be$, $^{10}B$, and
$^{11}B$ are thought to be produced in the interstellar medium
 by cosmic
ray spallation and fusion reactions
of protons and alpha particles with C, N, O and
He nuclei. Their produced abundances in the standard big bang
nucleosynthesis
 model is four to five orders of magnitude
below the observations (Kajino \& Boyd, 1990; Thomas et al., 1993a).
Production by cosmic rays can by and
large account for the observed
solar abundances of these elements (Reeves et al. 1970; Meneguzzi
et al.
1971). However, $^7Li$ is also partly produced by galactic
cosmic-ray nucleosynthesis (GCRN)
providing a potential constraint on big bang nucleosynthesis
(Olive \& Schramm 1992).
Because ratios of the isotopes produced in GCRN are largely model
independent,
observations of the $Be$ and $B$ abundances can be an important
test in determining what fraction of $^{7}Li$  observed in warm
halo subdwarfs
is primordial.  Conventional
GCRN models for the production of these elements
which assume time-independent spectra and follow the metallicity
of the ISM
 were able to explain the limited amount of data available at that time
(i.e. for $[Fe/H] \ge -1.4$) (Vangioni-Flam et al. 1990).  However
these simple models tend
to underproduce Be and B at low metallicities. These elements have now
recently been observed in low metallicity population II stars
(Rebolo et al.
1988; Ryan et al. 1990; Gilmore, Edvardsson \& Nissen 1991;
Ryan et al. 1992;
Gilmore et al. 1992;
 Duncan, Lambert \& Lemke 1992; Rebolo et al. 1993).
More recent models which treat the early galaxy as a ``closed"
rather than a
``leaky" box for GCR increase the predicted abundances of Be
and B for low
metallicities almost to observed levels (Prantzos, Cass\'{e},
\& Vangioni-Flam 1992, hereafter PCV).

It should be noted that
inhomogeneities present in the early universe may alter the predictions
of standard big bang nucleosynthesis (Applegate, Hogan \& Scherrer 1987;
 Alcock, Fuller \& Mathews 1987)
and give rise to increased nucleosynthesis beyond A = 8
 ( Boyd \& Kajino 1989;
Malaney and Fowler
1989). However it has been shown (Thomas et al 1993b) that even
under extreme conditions
the abundances of $^6Li$, $Be$, and $B$ in inhomogeneous big bang
nucleosynthesis
remain two to three orders of magnitude
below their observed values in halo dwarf stars.
Furthermore when the constraints from the light elements are applied,
there
is little distinction between standard and inhomogeneous model
results for elements
with $A < 12$ (Kurki-Suonio
et al. 1990; Thomas et al. 1993b; Reeves 1993).

While the abundances of Li, Be, and B predicted by the improved
models of GCR production of PCV almost match observations,
these models
cannot solve the  long-standing problem of the solar $^{11}B/^{10}B$
ratio.  Most GCR models
predict a ratio of around 2.5 (Meneguzzi et al. 1971; Meneguzzi and
Reeves
1975; Walker et al. 1985; Abia and Canal 1988; Steigman and Walker 1992;
Walker et al. 1993).
The observed ratio is found to be closer to 4
(Shima 1963; see also, Cameron 1982, Anders and Grevesse 1989).
Meneguzzi
and Reeves (1975) suggest as a possible solution to this problem
introducing a steeply decreasing spectral component at low energies
(less
than $80$ MeV/n) in GCR.  At these energies, the cross sections of $^{10}B$
and $^{11}B$ are somewhat different, leading to a larger isotopic ratio.
PCV point out, however, that $\alpha + \alpha \rightarrow~ ^{6,7}Li$
reactions
were ignored in these studies.  PCV find that steepening the GCR
spectra of protons and alphas at low energies leads to a solar isotopic
ratio of $^{11}B/^{10}B \sim 4$, but results in
$Li$ production considerably exceeding observed values in halo dwarfs.

PCV also explore a possible solution to this problem
by assuming that the more fragile
$^{10}B$ is completely
destroyed when ejected by a star of any size, while all of
the $^{11}B$ is
allowed to survive. Most previous models assume that all of
the Li, Be, and
B is destroyed when ejected from stars.
The maximum isotopic ratio obtained under these
conditions is 2.9.  They conclude that this can not account for the
observed isotopic ratio.

With the difficulty in producing the observed isotopic ratio of
$^{11}B/^{10}B$, it has been suggested that alternative
astrophysical sites
for the production of $^{11}B$ must be found.
One such site is at the shock front of type II supernovae, as
suggested by
Dearborn et al. (1989):  when the shock hits the
hydrogen envelope, it burns the ambient $^3He$ and $^4He$
producing $^7Be$.
Some of the resulting
$^7Be$ combines with alpha particles to produce $^{11}C$ which decays to
$^{11}B$.  The primary goal of that work was to explore an alternative
site for the production of $^7Li$ to reach Pop I abundances.  They
noted however that significant $^{11}B$ production might also take place.
To date, the $^{11}B$ produced by this process
has not been included in chemical evolution models and an isotopic ratio
of $^{11}B/^{10}B$ including production of $^{11}B$
by this process has not been
determined.

A potentially more important source for $^{11}B$ production has been
found to result from neutrino
induced nucleosynthesis in type II supernovae (Woosley et al. 1990).
The core
collapse of a massive star into a neutron star creates a flux of
neutrinos
so great that in spite of the small cross sections involved, it may
still induce substantial
nucleosynthesis.  It is found that considerable $^{11}B$ production
can result as the
flux of neutrinos passes through the He, C, and Si shells of the
stellar envelope, primarily by neutrino
spallation of $^{12}C$.  In addition, some synthesis of $^7Li$
and $^{10}B$ takes
place by this process but the production rate seems quite low.

In this paper, we include
production of $^{11}B$, $^7Li$, and $^{10}B$ predicted from
neutrino nucleosynthesis in type II supernova events
(Woosley et al. 1993; Timmes et al. 1993)
in galactic chemical evolution models along with GCR spallation
production.  We investigate the isotopic ratio of $^{11}B/^{10}B$
resulting from these models as well as the observational
consequences of the
increased abundances of $B$ and $Li$.
As we will show, although it is possible to obtain a
$^{11}B/^{10}B$ isotopic ratio $\simeq 4$, the $B/Be$ ratio
is typically high,
$B/Be \ga 20$ and is only marginally consistent with present
data for halo stars.
Indeed, this primary source of $^{11}B$ predicts significantly higher
$B/Be$ ratio at low metallicity than in standard models of GCRN
and can as such conclusively test this hypothesis.

The importance of the $\nu$-process was previously considered
 by Malaney (1992)
primarily with regard to the production of $^9Be$, though a
discussion of the effect on
$B$ and the $B/Be$ ratio was included. Malaney claimed that
 $^9Be$ is produced via
the $^7Li(t,n)^9Be$ reaction.  This rate was included in the recent
Timmes et al. (1993) calculations and no substantial amount of $^9Be$
was found to survive the supernova shock.

\section{Galactic Evolution from the $\nu$-process}

Before integrating the $\nu$-process yields into a full-blown model
of galactic chemical evolution, we make the following, more
simplified evaluation. Because we include a new source of $B$
without any accompanying $Be$, we must be sensitive to the
potential danger of an excessive $B/Be$ ratio.  In fact, this
ratio has the possibility of strongly constraining the $\nu$-process
source for $B$.  The $Be$ abundance observed in halo dwarf stars
shows an almost linear dependence with metallicity.  The observed $Be$
abundance can be fit approximately by
\begin{equation}
[Be] \equiv 12 + \log (Be/H) \simeq 1.7 \pm 0.4 + [Fe/H]
\end{equation}
Let us consider first the very simple model used in Steigman and
Walker (1992) and Walker et al. (1993).  There, it was assumed that
the Pop II abundances of $C$,$N$ and $O$ in ISM were simply related
to the
iron abundance as $[C/H] = [N/H] = [Fe/H]$ and $[O/Fe] =0.5$.
Evolution of
these elements was not included in the model. Furthermore, it was also
assumed that all exposure times for GCRN processes were equal.
The overall exposure time then can be fixed by fitting the $Be$
observations. Because the $B/Be$ ratio in this model is relatively
fixed at $B/Be \sim 14$ (i.e. the ratio of the corresponding
spallation cross-sections),  we can predict the $B$ abundance as a
function of $[Fe]$, $\log (B/H) \sim -9.2 + [Fe/H]$ which is in good
agreement with the fit to the three observations of $B$ when a linear
relation is {\it assumed},
\begin{equation}
[B] \equiv 12 + \log (B/H)  \simeq 2.7 \pm 0.9 + [Fe/H]
\end{equation}

The basis for the additional source of $^{11}B$ are neutrino
processes occurring during type II supernovae.  Neutrinos, including
also $\nu_\mu$ and $\nu_\tau$, are copiously produced in the hot
collapsed core during a supernova (see eg., Mayle, Wilson and
Schramm 1987).  Because of their higher temperature, $\nu_\mu$
and $\nu_\tau$ neutral current reactions are dominant. The inelastic
scattering of these neutrinos leads to unstable excited states which
decay by p,n or $\alpha$ emission.  These processes were included
in supernova nucleosynthesis calculations by Woosley et al. (1990)
where it was found that significant amounts of $^7Li$ and $^{11}B$
and lesser amounts of $^{10}B$ were produced (as well as other
heavier isotopes).

An important aspect of the calculation of Woosley et al. (1990) is the
full treatment of pre- and post-shock nucleosynthesis.
Since the duration of the neutrino burst exceeds the time scale for
the passage of the shock through the inner layers of the exploding
star, $\nu$-process nucleosynthesis can continue after the passage of
the shock. In the outer layers, however, the destruction of fragile
isotopes is a significant effect and is, for example, responsible
for the
destruction of $^9Be$. Given the nature of the calculation
there is a fair amount of uncertainty regarding the yields obtained
by this process.
The main uncertainty comes from the poorly known neutrino
spectrum; indeed, there are indications that the high energy part of it
(the most efficient for nucleosynthesis) should be considerably
supressed (Myra \& Burrows 1990).
In this work, we include the recent results of Timmes et al.
(1993) regarding the production of $^7Li$, $^{10}B$, and $^{11}B$
over a wide range of initial stellar masses. However, for the purposes
of normalization to a solar $^{11}B/^{10}B$ ratio we take the liberty
of adjusting the neutrino-process yields by an overall factor.

To estimate the effect of the $\nu$-process on the boron isotopic
ratio, we first separately integrate the $\nu$-process yields over
a simple initial mass function with slope parameter, $x = 1.8$, and a
constant star formation rate.
We find as expected an approximately linear relationship for $^{11}B$
versus
metallicity, $[B] = 2.55 + [Fe/H]$. When added to the GCRN component for boron,
 we find that the $^{11}B/^{10}B$ isotopic ratio is
raised from the canonical GCRN value of $\sim$ 2.5  to 4.5. Furthermore,
the $B/Be$ ratio is increased from $14$ to  $\simeq 22$. When
compared with the
available data: $[B] = 0.4 \pm 0.2$ (HD 19445); $= -0.1 \pm 0.2$
(HD 140283); and
$= 1.7 \pm 0.4$ (HD 201891) from Duncan et al. (1992) and
$[Be] < -0.3$ (HD 19945)
from Ryan et al. (1990); $[Be] = -1.25 \pm 0.4$ (HD 140283)
from Ryan et al. (1992);
$= -0.97 \pm 0.25$ (HD 140283) from Gilmore et al. (1992);
and $= 0.4 \pm 0.4$ (HD 201891)
from Rebolo et al. (1988). Thus the $B/Be$ ratio becomes
(propagating the quoted errors):
$B/Be = 20 \pm 26$  (HD 201891); $B/Be = 14 \pm 14$
(HD 140283)  based on the beryllium measurement of Ryan et al. (1992)
and $B/Be = 7 \pm 5$ (HD 140283) based on the measurements of
 Gilmore et al. (1992);
and $B/Be > 5$ (HD 19445)\footnote{
to date only an upper limit for Be is available for this star
(Ryan et al. (1990)},
one sees that although the agreement with
the data is worsened when the $\nu$-process yield is included,
 this explanation
of the boron isotopic ratio can by no means be excluded.

Notice that in this section we assume a GCR origin for $Be$,
$^{10}B$ and (part of)
$^{11}B$, but we adopt the {\em observed} $B$ vs. $Fe$ and
$Be$ vs. $Fe$ relationships
which are almost linear.  It is well known however, that such
linearity characterizes the
evolution of a primary element (like oxygen), whereas in
conventional GCR models
$B$ and $Be$ behave as secondaries (see Prantzos et al. 1993b).
This discrepancy between
theory and the observations is somewhat alleviated, but not
completely removed, in PCV,
where a more efficient production of $B$ and $Be$ in the early
Galaxy is suggested (see
next section).  Obviously, if the $\nu$-process $B$ is
incorporated in such a model,
the resulting $B/Be$ ratio at low metallicity is expected to
 be even larger than
previously estimated. In the next section we investigate this
effect with a fully
developed model of galactic chemical evolution.

\section{$LiBeB$ production by neutrinos and cosmic rays}

We start with a brief description of the PCV model, where an important
correction has been incorporated.
It is currently thought that GCR are accelerated mainly by supernovae
shock waves and propagate
in the Galaxy by diffusing on the irregularities of the
galactic magnetic field
and suffering losses by ionization, nuclear reactions and
 leakage,  in the
framework of the ``leaky box" model (e.g. Cesarsky 1980).
Notice that in this
model leakage takes place perpendicularly to the galactic
disk, which has
today a confinement volume with thickness $H_0 \sim$0.5 kpc
and a gas fraction
$\sigma_{gas,0} \sim$ 0.10 - 0.20 (Rana 1991).
The corresponding {\it escape length} (a key parameter
caracterizing the leaky box
model and indicating the amount of matter that GCR
``traverse" before being lost)
is estimated to be $\Lambda_{e,0} \sim$ 10 g/cm$^2$, at
 least up to 3-4 GeV/nucleon.

In order to explain the recent B and Be observations in halo stars,
PCV suggest that during its
early evolution the Galaxy was a ``closed" rather than
a ``leaky" box to GCR.
According to the PCV analysis, the escape length
evolves as $\Lambda_e \propto
\ \sigma_{gas} \ H$, i.e. it was much larger in the
past due to the larger
 gas content and the larger dimensions of the young
Milky Way. The corresponding
GCR spectrum is found to be flatter than the current
one, and the associated
proton flux above $\sim$100 MeV is found to evolve
roughly as $F \propto \sqrt \Lambda_e$.
Using a simplified prescription for the early
collapse phase of the Galaxy and
a detailed chemical evolution model, the recent
Be and B observations
are well reproduced with this hypothesis on
GCR propagation. Notice, however,
that the obtained B vs. Fe and Be vs. Fe relationships
are still non linear,
i.e. B and Be are clearly produced as secondary elements in that model.

Since that work, Malaney and Butler (1993) noticed that
the loss of protons due to
$\pi^o$ production has been neglected in PCV. This term is insignificant
today. Indeed, the {\it nuclear destruction length} for
protons is $\Lambda_N
\sim$ 200 gr cm$^{-2}$ and does not affect the total loss length
$\Lambda^{-1} = \Lambda_N^{-1}  + \Lambda_e^{-1} \sim \Lambda_e^{-1}$.
However,   this factor
might have been important in the earliest phases of the Galaxy,
{\it if GCRs were
as efficiently confined as suggested in PCV}. In this work we
incorporate the
nuclear destruction term in our model. As a result we
 obtain GCR fluxes lower by a
factor of $\sim$2 w.r.t. the previous work at large $\Lambda$  values
(see Fig. 1 in Prantzos and Cass\'e  1993). In the framework
of the galactic evolution model this concerns quite a
short period (a few 10$^7$ years), in the very beginning
of the Galaxy. The
resulting Be and B abundances are slightly affected, but still
consistent with the observations, as can be seen in
Fig. 1 and 2. Notice also
that the nuclear destruction term is much more important
for alpha particles,
which have a corresponding length of
$\Lambda_N \sim$ 25 gr cm $^{-2}$  only.
For that reason, GCR alphas never develop
spectra as flat as GCR protons,
even in the case of a good confinement: they are simply
``removed" by collisions
with the ISM protons for values of $\Lambda \geq$ 25 gr cm$^{-2}$.
In this modified scenario (i.e. maintaining the efficient
GCR confinement
in the early Galaxy, but incorporating the nuclear destruction term)
the early Li abundance is maintained low and consistent with
 observations (see Fig.3)
The reason, however, is not the flattening of the GCR
alpha spectra (as in the
original PCV), but the nuclear destruction of alphas at large $\Lambda$.

Given the reasonable success of the modified PCV model at
 explaining the low
metallicity observations of $Be$ and $B$, we now include
new isotopic yields of Woosley et al. (1993)
and Timmes et al. (1993) of
$\nu$-process produced $^7Li$, $^{10}B$, and $^{11}B$. Our goal
is to explain of the $^{11}B/^{10}B$ ratio which is
observed to be $\sim 4$
in the solar system.
The PCV model, like almost all models of GCRN, gives a
ratio of about $2.5$. The additional
source of $^{11}B$ from the $\nu$-process increases
 this ratio. We have adjusted the overall
$\nu$-process yield to give a solar system
 $^{11}B/^{10}B$ ratio of 4. To do so, we found
that it was necessary to reduce the neutrino induced
yields by a factor $\sim 2$ to avoid overproducing $^{11}B$.
A summary of our results is displayed in Table 1.
The resulting  $^{11}B/^{10}B$ ratio as a function of
metallicity is shown in Figure 4.
This result can be compared directly to the case without
 a $\nu$-process contribution
(dotted line) and to the case without an enhanced
 cosmic-ray spectrum (dashed line).
As one can see, enhancing the cosmic-ray spectrum
 has little effect on the $^{11}$B/$^{10}$B
ratio.
The large discontinuity is solely due to our
treatment of the two (halo and disk)
separate components of the galaxy. (See PCV for details of the model.)
A one zone model and presumably a more comprehensive
treatment of the two zones (including infall in the disk)
 would show a smoother relationship.

With the $^{11}B/^{10}B$ ratio fixed, we can now examine
 the compatibility of the model
with the data on $Li$, $Be$, and $B$.  In Figures 1-3,
 we show the evolution of the $LiBeB$
isotopes as a function of $[Fe/H]$. In each of these figures,
we compare the abundance of the $LiBeB$ element with (solid)
and without (dotted)
$\nu$-process nucleosynthesis and without an enhanced
cosmic-ray spectrum (dashed).
The solar values for $B$ and $Be$ are the same as those adopted in PVC.  They
are somewhat
smaller than the values given by Anders \& Grevesse (1989)
and are closer to the values
adopted in Reeves and Meyer (1978) and Arnould \& Forestini (1989).
In Figure 1, we show the beryllium abundance as a function of $[Fe/H]$.
  As discussed above, the beryllium is slighlty
lower than the results of PCV due to the correction in the
escape length noticed by
Malaney and Butler (1993), but is never-the-less still in
acceptable agreement with the
data. The $\nu$-process makes very little contribution to $Be/H$.
 As shown in Figure 2, the model appears acceptable with
respect to the evolution
of the boron abundance both with and without the $\nu$-process
 contribution. For both
beryllium and boron, the benefit of the enhanced cosmic-ray
 spectrum is clear. Finally
we see that the lithium abundance shown in Figure 3 is also
 quite compatible with the
data. Here, we have taken a primordial abundance
 $[Li] \simeq 2.0$ (Olive and Schramm
1992) and have included both the GCRN and
$\nu$-process contributions, neither of which
is excessive at low metallicities. (Notice
that since no stellar source of $Li$ is included
in the model, the predictions for the disk
($-1 < [Fe/H] < 0$) are lower that the
observations.)

Despite the apparent success regarding the observed
 abundances of the $LiBeB$ elements,
the $B/Be$ ratio is  more difficult to reproduce.
 In Figure 5, we show the $B/Be$
ratio as a function of $[Fe/H]$.  Though at
 $[Fe/H] \simeq -1$ to $-2$,
the ratio is only slightly higher than
 standard GCRN predictions and consistent with
the data for HD 201891, at lower
$[Fe/H] < -2$ there is a clear departure
from more standard model results.
The $B/Be$ becomes very large ($B/Be > 50$), for
$[Fe/H] \la -3$. This is clearly due
to the primary nature of $^{11}B$ produced mostly
by the $\nu$-process in the halo vs.
the secondary nature of $Be$,
produced only by GCRN.  As it stands,
the model is only marginally consistent with the
data for HD 140283 (at the 2-$\sigma$ level).
 With regard to HD 140283, we note the following:
1) A recent observation (Molaro et al. 1993)
has failed to identify beryllium in this star
using a potentially more reliable part of the
line spectrum and indicate that
a lower value for $Be/H$ and hence higher value for $B/Be$ is likely;
2) The oxygen abundance in this star seems
perhaps anomalously high.  Several
measurements indicate a value $[O/Fe] \sim 0.8 - 0.9$
 (Bessel and Norris 1987;
Molaro et al. 1993). Such a large oxygen enhancement
 may also be accompanied by a
beryllium enhancement due to cosmic-ray spallation
(the boron enhancement would be masked by the
$\nu$-process contribution). Thus this star seems
 particularly incapable
of excluding the high value for $B/Be$ we would
expect at $[Fe/H] \sim -2.6$
due to the $\nu$-process enhancement of $^{11}B$.
 Future measurements of $B$ and $Be$ at
low metallicity have the clear potential at
confirming or rejecting the $\nu$-process
contribution as a solution to the $^{11}B/^{10}B$ isotopic ratio.

\section{Conclusion}

In summary, we have tested the hypothesis that a primary
source of $^{11}B$ due to
$\nu$-process nucleosynthesis in type II supernovae can explain
the observed solar $^{11}B/^{10}B$ isotopic ratio.
We note at this point that the observed value
$^{11}B/^{10}B \simeq 4$ is due to a relatively old
measurement (Shima 1963).
Given the attention paid to this number, a new
 measurement of this ratio is certainly warranted.
Though we are able to produce the oberved isotopic
ratio of 4 as well as a reasonable
evolution for the $LiBeB$ elements, we expect a
significantly higher $B/Be$ ratio
than in more standard models of cosmic-ray
nucleosynthesis and galactic chemical
evolution. Due to the primary nature of the
neutrino-process source for $^{11}B$
relative to the secondary source for $Be$,
we have found that the ratio
$B/Be > 50$ when $[Fe/H] < -3$.  Indeed, we
think that measurements of the $B/Be$
ratio in halo stars can help clarify the
importance of neutrino-process nucleosynthesis
(and ultimately confirm or invalidate it).

}
}
\newpage
\noindent {\bf Acknowledgements}
\vskip 0.3truecm
We would like to thank Frank Timmes and Stan Woosley for
generously sharing
their most recent results of supernova nucleosynthesis.
We would also like to thank
Jean Audouze, Michel Cass\'{e}, and David Schramm for
useful discussions.
KAO would like to thank the Institut d'Astrophysique de Paris where
this work was completed for their hospitality.
The work of KAO and SS was supported in part by
DOE grant DE-AC02-83ER-40105.
The work of KAO was in addition supported by a Presidential Young
Investigator Award. The work of NP and EV-F was
 supported in part by PICS $n^o$114,
``Origin and Evolution of the Light Elements", CNRS.
\vskip 1in
\begin{table}[h]
\begin{center} {\sc Table 1.  Comparison between observed
values at the birth of
the solar system at $t = 12.5$ Gyr and the model}
\end{center}
\begin{center}
\begin{tabular}{lcc}
 \hline \hline
                    & Observed Data           & Model Result \\ \hline
$X_H/X_{H_\odot}$                  & 1         &   1.03    \\
$Y/Y_\odot$                        &1          &    0.92   \\
$X_{Be}/X_{{Be}_\odot}$            & 1          &   1 (normalization)    \\
$X_{B}/X_{{B}_\odot}$            & 1           &    1.08   \\
$^{11}B/^{10}B$             & $4.05 \pm 0.1$           &
   4.11 ($\nu$-process yield adjusted)   \\
$[O/H]$             & 0               &   0.13   \\
$[Fe/H]$               &0                &0.011   \\
$Z/Z_\odot$             & 1           &   0.94   \\
$\sigma$ (present value)              & 0.10 - 0.20
   &   0.19   \\ \hline
\end{tabular}

\end{center}
\end{table}
\newpage
\beginapjbib
\bibitem Abia, C., \& Canal, R. 1988, A \& A, 189, 55

\bibitem Alcock, C., Fuller, G.M., \& Mathews, G.J.
1987, ApJ, 320, 439

\bibitem Anders, E., \& Grevesse, N. 1989, Geochim.
Cosmochim. Acta, 53, 197

\bibitem Applegate, J.H., Hogan, C.J., \& Scherrer, R.
1987, Phys. Rev., D35, 1151

\bibitem Arnould, M. \& Forestini,M. 1989, Nuclear
Astrophysics, ed. M. Lozano et al.
(Berlin:Springer), 48

\bibitem Bessel, M.S., \& Norris, J.E. 1987, Jour.
Astrophys. \& Astron. 8, 199

\bibitem Boyd, R.N., \& Kajino, T. 1989, ApJ, 336, L55

\bibitem Cameron, A.G.W. 1982, in {\it Essays in
 Nuclear Astrophysics}, ed. C.
Barnes, D. Clayton, and D. Schramm (Cambridge:Cambridge
University Press), p.35

\bibitem Cesarsky, C. 1980, ARA\&A, 18, 289

\bibitem Dearborn, D. S. P., Schramm, D., Steigman, G. \& Truran, J.
1989 ApJ, 347, 455

S. D. 1990 ApJS, 73, 21


\bibitem Duncan, D. K., Lambert, D. L., \& Lemke,
M. 1992 ApJ, submitted

\bibitem Gilmore,G., Edvardsson,B. \& Nissen,
P.E. 1991 Astrophys. J. 378, 17.

\bibitem Gilmore,G., Gustafsson,B., Edvardsson,B.
\& Nissen,P.E.  1992, Nature,
357, 379



\bibitem Kajino, T., \& Boyd, R.N. 1990, ApJ, 259, 267

\bibitem Kurki-Suonio, H., \& Matzner, R.A., Olive, K.A., \&
	Schramm, D.N. 1990, ApJ, 353,406

\bibitem Malaney, R.A. 1992, ApJ, 398, L45

\bibitem Malaney, R.A., \& Butler, M.N. 1993 ApJ, 407, L73

\bibitem Malaney, R.A., \& Fowler, W.A. 1989, ApJ, 345, L5

\bibitem Mayle, R.W., Wilson, J.R., \& Schramm, D. N. 1987,
ApJ, 318, 288

\bibitem Meneguzzi, M., Audouze, J.,\& Reeves, H.
1971 Astron. \& Astrophys., 15,337

\bibitem Meneguzzi, M., \& Reeves, H. 1975 Astron.
 \& Astrophys., 40,110

\bibitem Molaro, P., Castelli, F., \& Pasquini, L. 1993,
 in {\it Origin
 and Evolution of the Light Elements} eds. N. Prantzos,
E. Vangioni-Flam, and M. Cass\'{e}
(Cambridge:Cambridge University Press), p. 153

\bibitem Myra, E., \& Burrows, A. 1990, ApJ, 364, 222

\bibitem Olive, K. A., \& Schramm, D. N. 1992 Nature, 360,439


\bibitem Prantzos, N., \& Cass\'{e}, M. 1993, ApJ, submitted

\bibitem Prantzos, N., Cass\'{e}, M., \& Vangioni-Flam, E.
 1993a ApJ, 403, 630 (PCV)

\bibitem Prantzos, N., Cass\'{e}, M., \& Vangioni-Flam, E.
1993b in {\it Origin
 and Evolution of the Light Elements} eds. N. Prantzos, E.
Vangioni-Flam, and M. Cass\'{e}
(Cambridge:Cambridge University Press), p. 156

\bibitem Rana, N. 1991, ARA\&A, 29, 129

\bibitem Rebolo, R., Molaro, P., Abia, C. \& Beckman, J.E. 1988
 Astron. Astrophys. 193, 193


\bibitem Rebolo, R., Garcia Lopez, R. J., Martin, E. L.,
 Beckman, J. E.,
McKeith, C. D., Webb, J. K., \& Pavlenko, Y. V. 1993 in {\it Origin
 and Evolution of the Light Elements} eds. N. Prantzos,
 E. Vangioni-Flam, and M. Cass\'{e}
(Cambridge:Cambridge University Press), p. 149

\bibitem Reeves, H. 1993, in {\it Origin
 and Evolution of the Light Elements} eds. N. Prantzos, E.
 Vangioni-Flam, and M. Cass\'{e}
(Cambridge:Cambridge University Press), p. 168

\bibitem Reeves, H. Fowler, W.A. \& Hoyle, F. 1970
{\sl Nature}, 226, 727

\bibitem Reeves, H. \& Meyer, J.P. 1978, ApJ, 226, 613

\bibitem Ryan, S.,  Bessel, M., Sutherland, R.,
\& Norris, J. 1990 ApJ, 348, L57

\bibitem Ryan, S., Norris, J., Bessel, M.,
\& Deliyannis, C. 1992 ApJ, 388, 184

\bibitem Shima, M. 1963,Geochim. Cosmochim. Acta , 27, 911.





\bibitem Spite, F., \& Spite, M. 1991 A\&A, 252, 689

\bibitem Steigman, G., \& Walker, T. P. 1992 ApJ, 385, L13

\bibitem Thomas, D., Schramm, D. N., Olive, K. A., \& Fields, B. D.
1993a ApJ, in press

\bibitem Thomas, D., Schramm, D. N., Olive, K. A.,
Meyer, B., Mathews, G. J., \& Fields, B. D. 1993b (in preparation)

\bibitem Timmes, F. X., Woosley, S. E., \& Weaver, T. A.
1993 ApJ (in preparation)

\bibitem Vangioni-Flam, E.,  Cass\'{e}, M., Audouze, J., \& Oberto, Y. 1990,
ApJ, 364, 568

\bibitem Walker, T.P., Mathews, G.J.\& Viola, V.E. 1985 ApJ 299, 745

\bibitem Walker, T. P., Steigman, G., Schramm, D. N., Olive, K. A.,
\& Fields, B. 1993 ApJ, in press

\bibitem Walker, T. P., Steigman, G., Schramm, D. N., Olive, K. A.,
\& Kang, H. 1991 ApJ, 376, 51

\bibitem Woosley, S. E., Hartmann, D.H., Hoffman, R.D. \& Haxton, W. C. 1990
ApJ, 356, 272

\bibitem Woosley, S. E., Timmes, F. X., \& Weaver, T. A
1993, in {\it Nuclei in the Cosmos},
eds. F. K\"appeler \& K. Wisshak, (London:Institute of
 Physics Publishing)

\endapjbib

\newpage
\noindent{\bf{Figure Captions}}

\vskip.3truein

\begin{itemize}

\item[]
\begin{enumerate}
\item[]
\begin{enumerate}
\item[{\bf Figure 1:}] The $Be$ abundance as a function
of $[Fe/H]$ with (solid) and without (dotted)
$\nu$-process nucleosynthesis included.  The two curves coincide
since $^9Be$ is not produced by the $\nu$-process.  Also shown
 is the $Be$ abundance
in an unenhanced GCRN model (dashed).  The theoretical curves are
compared to available data. (See text for references on
 observational data.)

\item[{\bf Figure 2:}] The $B$ abundance as in Figure 1.

\item[{\bf  Figure 3:}] The $Li$ abundance as in Figure 1.
 Data are from
Spite \& Spite (1991).  Only the upper envelope of the data is shown.

\item[{\bf  Figure 4:}]  The $^{11}B/^{10}B$ ratio as in Figure 1.

\item[{\bf  Figure 5:}]   The $B/Be$ ratio as in Figure 1.

\end{enumerate}
\end{enumerate}
\end{itemize}

\end{document}